\documentclass[default,iicol]{sn-jnl}

\usepackage{amsmath}
\usepackage{amssymb}
\usepackage{graphicx}

\usepackage{natbib}
\usepackage{hyperref}
\usepackage{doi}

\begin{document}


\title[Article Title]{Elasto-frictional reduced model of a cyclically sheared container filled with particles}

\author[1,2]{\fnm{Antoine} \sur{Faulconnier}}
\author*[1]{\fnm{Stéphane} \sur{Job}}\email{stephane.job@isae-supmeca.fr}
\author[2]{\fnm{Julien} \sur{Brocail}}
\author[1]{\fnm{Nicolas} \sur{Peyret}}
\author[1]{\fnm{Jean-Luc} \sur{Dion}}

\affil[1]{\orgdiv{Laboratoire EULER}, \orgname{ISAE-Supméca}, \orgaddress{\street{3 rue Fernand Hainaut}, \city{Saint-Ouen-sur-Seine}, \postcode{93400}, \country{France}}}

\affil[2]{\orgdiv{Estaca’Lab}, \orgname{Estaca campus Ouest}, \orgaddress{\street{rue Georges Charpak}, \city{Laval}, \postcode{53000}, \country{France}}}


\abstract{ 
%
This article explores the hysteretic behavior and the damping features of sheared granular media using discrete element method (DEM) simulations. We consider polydisperse non-cohesive frictional spherical particles, enclosed in a container with rigid but moving walls, subjected to a cyclic simple shear superimposed on a confining pressure. The mechanical response of the grains is analyzed in the permanent regime, by fitting the macroscopic stress-strain relation applied to the box with a Dahl-like elasto-frictional model. The influence of several parameters such as the amplitude of the strain, the confining pressure, the elasticity, the friction coefficient, the size and the number of particles are explored. We find that the fitted parameters of our macroscopic Ansatz rely qualitatively on both a well-established effective medium theory of confined granular media and a well-documented rheology of granular flow. Quantitatively, we demonstrate that the single degree-of-freedom elasto-frictional reduced model reliably describes the nonlinear response of the granular layer over a wide range of operating conditions. In particular, we show that the mechanical response of a granular slab under simple shear depends on a unique dimensionless parameter akin to an effective Coulomb threshold at low shear/high pressure. Furthermore, exploring higher shear/lower pressure, we evidence optimal damping at the crossover between a loose unjammed regime and a dense elastic regime.\\
\begin{center}
\textbf{Date}\\
September 17, 2025
\end{center}
} 

\maketitle


\section{Introduction}

Granular materials are ubiquitous in Nature, where they are found in a variety of states of matter~\citep{Jaeger1996}.
From the application point of view, granular media are known to dissipate and disperse mechanical shocks efficiently ({\em a sand pile can stop a rifle bullet}) and they have also proved to be efficient in the context of sound and vibration mitigation~\citep{Melo2006}, damping~\citep{Masmoudi2016} and insulation~\citep{Hentati2021}.
Compared to classical matter, granular media present unusual properties like softening~\citep{Johnson2005}, weakening~\citep{Jia2011}, phase transition~\citep{Zuniga2022}, relaxation~\citep{Ren2013} and aging~\citep{Espindola2016}, to name but a few.

Jammed granular media are frequently referred to as granular solids owing to their ability to resist pressure and shear, or even as metamaterials~\citep{Gotz2023} to tailor wave absorption~\citep{Vasconcellos2022} and propagation~\citep{Vasconcellos2024}. 
To determine their elastic features, an effective medium theory (EMT) was developed in the late 1980s to connect elastic moduli at the particle level to that at the sample scale~\citep{Walton1987}.
Such a macroscopic description of a pile of grains acquires and reveals the intrinsically nonlinear Hertzian potential at the heart of contact mechanics~\citep{Johnson1985,Popov2010}.
When the particles are non-cohesive and dry, they are also likely subjected to friction~\citep{Popov2010}.
Frictional properties are studied in particular in the field of granular flow, in line with many industrial applications, where one of the bottlenecks relies on understanding how friction dictates the rheology of an avalanche, for instance~\citep{gdrmidi2004}.
Here, a key description is found in dense and slow granular flows, where such behavior can be described with the tools and definitions of soil mechanics~\citep{Rognon2008,Jop2015,Fall2017}.
In this context, an effective friction coefficient is defined as the ratio between shear stress and confining pressure~\citep{gdrmidi2004}. 
Interestingly, this macroscopic coefficient depends only on a dimensionless inertial number $I$~\citep{Cruz2005,Jop2006}.

Inter-particle friction, together with inelastic collisions between grains, is also a key feature to damp mechanical energy in discrete systems. Indeed, vibrated granular media are able to achieve perfectly inelastic collapse conditions~\citep{McNamara1992,Pacheco2013}, leading to exceptional damping efficiency~\citep{Sanchez2012,Michon2013,Masmoudi2016,Ferreyra2021}.
Here, as dry friction is mostly a frequency independent mechanism, it is much more effective at low frequency than viscoelastic damping~\citep{Kausel2002,Rouleau2013}.
Phenomenologically, the friction between numerous particles gives rise to powerful hysteretic behaviors at the sample scale~\citep{Bi2011,Ren2013}, which themselves rely on multiple stick-slip events at the contact scale, leading in practice to nonlinear damping properties~\citep{Assimaki2000,Brandes2010,Senetakis2013,Awdi2023}.
The macroscopic hysteretic response is thus closely related to the anisotropy and the topological disorder of the contact network, as evidenced for instance by X-Ray tomography of granular media under shear~\citep{Xing2021}.

In the vast majority of the studies on granular media, simulations based on the discrete element method (DEM) are widely used to model soils~\citep{Khalili2017a,Khalili2017b} and structures made of particles~\citep{sitharam2010,Zhong2018} to explore in-depth phenomena at the microscopic scale; such a detailed description generally remains a challenging task in experiments, due to the large number of interacting bodies.
For instance, Luding {\em et al}~\citep{magnanimo2011,Krijgsman2013} used DEM simulations to build a constitutive model describing the behavior of a bi-axial box filled with particles subjected to pure shear in two-dimensions~\citep{magnanimo2011}, by taking into account various phenomena like anisotropy, hysteresis and ratcheting.
This model was later enriched to include pressure relaxation and stabilization in pure shear~\citep{Krijgsman2013}, making it possible to describe transient as well as permanent regimes. 
Other extensive models for granular matter include granular solid hydrodynamics~\citep{Jiang2009} handling static behavior with plasticity, as well as transient loading~\citep{Gudehus2019} including hysteresis.

From the practical engineering point of view, models were also derived to describe the macroscopic mechanical response of structural elements, such as a slender composite beam containing a granular medium core, from either simulations~\citep{Gotz2023} or experiments~\citep{Bajkowski2019}. In line with these attempts, our study aims at deriving a simplified but reliable description of a vibrated confined granular medium, capable of reducing the complexity of DEM simulations to facilitate, for instance, the optimization of vibration dampers containing particles~\citep{Fichant2024}. In particular, we require our reduced model to be parametrized by identifiable and measurable quantities, such as the features of the grains (size, material, etc.) and the features of mechanical action (confinement pressure, shear amplitude, etc.).
Starting from a DEM numerical model of individual grains confined in a box and subjected to cyclic simple shear, in this study we thus provide and analyze a reduced elasto-hysteretic model at the macroscopic scale.
The evolution of the elastic and the damping features of both the numerical model and the reduced model are explored, seeking in particular for configurations with a high damping ratio.

The paper is organized as follows. 
Sec.~\ref{sec:method} describes the setup of the DEM simulation to model a pack of grains enclosed in a box undergoing simple shear cyclically. 
In Sec.~\ref{sec:results}, the evolution of quantities of interest such as shear stress, pressure, elasticity and frictional behaviors are identified and examined at the light of an elasto-hysteretic Ansatz composed of a hysteretic term rendered by a Dahl model combined with a linear elastic response. The reliability of this model is probed via an extensive parametric study.
In Sec.~\ref{sec:discussion}, we evidence how our Ansatz relates to an effective medium theory describing the macroscopic elasticity of granular media, and to a rheological model describing granular flows.
Furthermore, we find through a dimensional analysis that the granular material can be described reasonably by a single non-dimensional parameter, akin to an effective Coulomb threshold.
The reliability of this model is discussed and then used to estimate the damping features. Finally, we propose a nonlinear improvement of the model, stemming from the anisotropy-like pressure fluctuations observed at high shear/low pressure; these fluctuations are described by a second non-dimensional parameter.
Sec.~\ref{sec:conclusion} provides the conclusion.


\section{Method and protocol}\label{sec:method}

We build a numerical version of a granular medium with a lateral size that is large enough compared to the average particle radius, to eliminate the discreetness of the medium at the scale of the representative elementary volume.
Considerations on the number, the size and the polydispersity of the particles are detailed in the Appendix Sec.~\ref{sec:A4_check_part_num_size}.
Our numerical sample thus consists of $1000$ ($ 10 \times 10 \times 10 $) polydisperse and purely elastic spherical particles (nominally made of steel: mass density $\rho_p=7850$~kg/m$^{3}$, Young's modulus $E_p=210$~GPa, Poisson ratio $\nu_p=0.27$, inter-particle friction coefficient $\mu_p=0.15$), with uniformly distributed radii to prevent crystallization ($r_p = 1\pm 0.25$~mm).
The grains are placed within a cubic box with rigid but movable walls.
The model is simulated using DEM with {\em Yade}~\citep{Smilauer2021} open source software.
To do this, Newton's equations of motion are integrated in the time domain with an explicit second order Velocity Verlet scheme, with a constant time step chosen equal to the time of flight of the longitudinal waves across the smallest particle, $\Delta t = \mathrm{min}(r_p/\sqrt{M_p/\rho_p})$, being $M_p=E_p(1-\nu_p)/(1+\nu_p)(1-2\nu_p)$ the P-wave modulus of particles material.
This time step is always smaller than the duration of inter-particle mechanical contact (i.e. the smallest duration in the system) because the deformation rate is smaller than the P-wave speed~\citep{Johnson1985,Popov2010}; this choice thus ensures a proper discretization in time to fulfill energy conservation.
Elasto-frictional interactions between particles are described by Hertz-Mindlin contact law coupled with a Mohr-Coulomb criterion, which means that the normal and the tangential contact forces $F_{n,t}$ versus deformations $\delta_{n,t}$ take the form~\citep{Johnson1985,Popov2010}
\begin{eqnarray}
F_n&=&\frac{2}{3}k_nr_p^{1/2}\delta_n^{3/2}\mbox{ with }k_n=\frac{2E_p}{1-\nu_p^2},\label{eq:Hertz_Mindlin_n}\\
F_t&=&k_t(r_p\delta_n)^{1/2}\delta_t\mbox{ with }k_t=\frac{8G_p}{2-\nu_p},\label{eq:Hertz_Mindlin_t}
\end{eqnarray}
if the particles overlap due to compression and stick due to friction; otherwise, both forces tend to zero when the non-cohesive particles do not overlap, $\delta_n\leq0$, and the tangential force saturates if sliding occurs, $F_t=\mathrm{min}(\|F_t\|,\mu_p F_n)\times\mathrm{sign}(\dot{\delta}_t)$. Here, $G_p=E_p/2(1+\nu_p)$ is the shear modulus of the particles’ material and $k_{n,t}$ are their normal and tangential contact stiffnesses.
Without loss of generality, the particle-to-wall interactions are chosen as frictionless ($\mu_w = 0$).
This choice implies that no shear stress occurs between the particles and the walls in any directions, and simplifies the interpretation on the origin of the dissipation which, in this case, only comes from inter-particle friction.

\begin{figure}[t]
\centering
\includegraphics[width=1\linewidth]{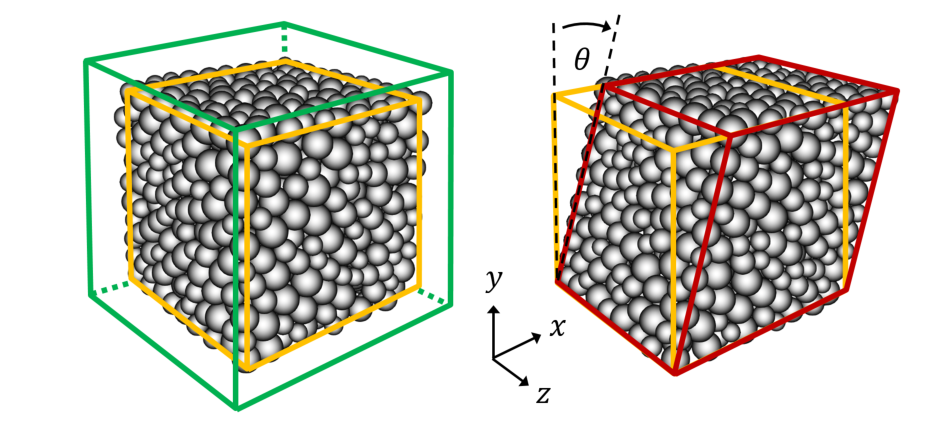} 
\includegraphics[width=1\linewidth]{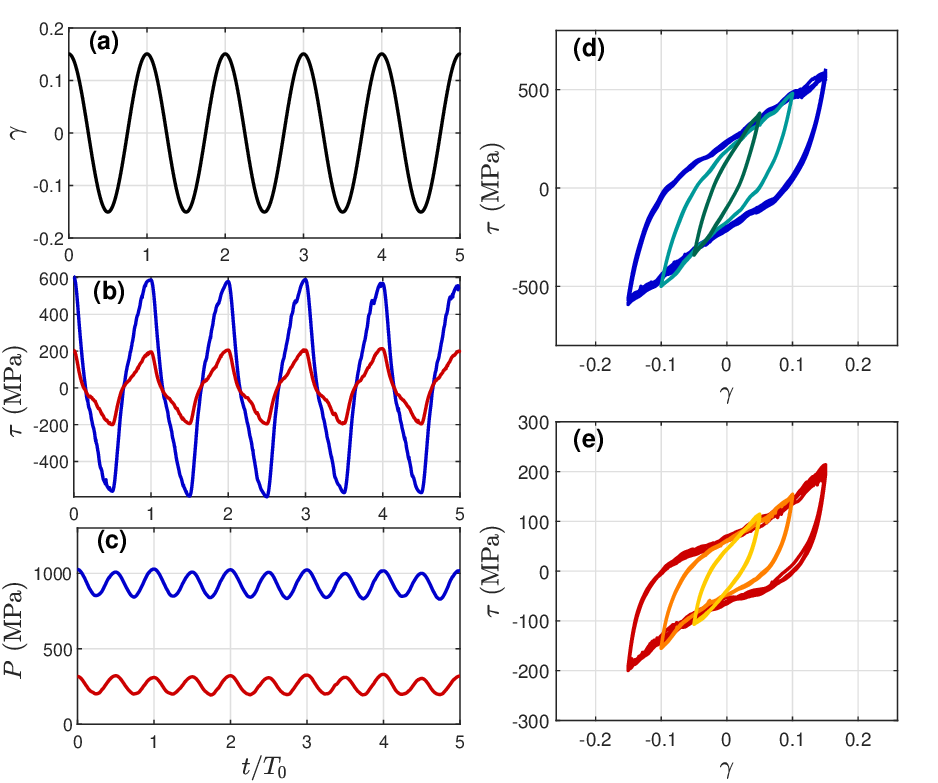}
\caption{
\textbf{(top)} Granular sample during \textbf{(left)} initial compression and \textbf{(right)} cyclic shear. \textbf{(a)} Imposed shear strain, \textbf{(b)} shear stress and \textbf{(c)} hydrostatic pressure, as functions of time for two values of the volume fraction $\phi = 0.64$ (red) and $\phi = 0.70$ (blue) at $\gamma_m=0.15$. Stress-strain curves for \textbf{(d)} $\phi = 0.70$ and \textbf{(e)} $\phi = 0.64$ at various amplitudes: $\gamma_m=0.05$, $0.10$ and $0.15$.}
\label{fig:1_sketchs_examples}
\end{figure}

In the {\em first step}, we prepare a granular sample. 
We generate the $1000$ size-distributed particles placed at random positions in a dilute state (at rest and well separated), inside a large cubic box delimited by walls, using a {\em Yade} builtin generator. 
Then, we set the sample under pressure, by proceeding with an isotropic compression until a given volume fraction $\phi = V_p/V$ is reached, with $V_p$ being the material volume of the spheres and $V$ the volume of the parallelepiped box. 
This step is achieved by imposing the displacement of all the walls at a constant and slow normal velocity, see Fig.~\ref{fig:1_sketchs_examples} (top-left); note the walls do not interact with each other.

In the {\em second step}, the granular packing is subjected to cyclic simple shear.
In practice, the box is deformed by imposing the rotation of two of the side walls, see the $yz$ planes in Fig.~\ref{fig:1_sketchs_examples} (top-right), similarly to what was done experimentally in~\citep{Xing2021}.
However, in our case, the volume of the box is kept constant, whereas the confining pressure is kept constant in~\citep{Xing2021}.
Thanks to the choice of a frictionless particle-to-wall interaction, the numerical model can be simplified by defining the four other walls, parallel to $xy$ or $xz$ planes in Fig.~\ref{fig:1_sketchs_examples}, as rigid and motionless. Otherwise, accounting for the parietal friction would require considering a deformable container to accommodate realistically the relative motion between grains and walls, which is beyond the aim of the present study and a simple DEM formulation.
We impose a harmonic rotation angle (in radians) $\theta(t)=\theta_m \sin(\omega_0 t)$ of the two $yz$ walls around the $z$-axis leading to a shear strain $\gamma = \tan(\theta)$ which is quasi-harmonic in the limit of small angles, $\gamma(t)\simeq\gamma_m\sin{(\omega_0t)}$ with $\gamma_m\simeq\theta_m\ll1$~rad.
It is important to note that, here and in the following, cyclic shear transformations are quasi-static in order to neglect inertia in the balance of forces and moments. 
The latter condition is $\omega_0 t_w\ll 1$ in practice, as the angular frequency is proportional to the shear rate, $\omega_0\propto\dot{\gamma} $, and the time of flight of a compressional wave across the granular medium is $t_w=r_p/\sqrt{M_\mathrm{eff}/\rho_\mathrm{eff}}$, with $\rho_\mathrm{eff}\propto\rho_p$ the mass density and $M_\mathrm{eff}=K_\mathrm{eff}+(4/3)G_\mathrm{eff}$ the longitudinal modulus of the sample's effective medium defined in Sec.~\ref{sec:EMT} below.

The DEM computation allows extracting the instantaneous quantities at the contact scale from which we compute the variables of interests and the state variables as functions of time: namely, the shear strain $\gamma(t)$, the shear stress $\tau(t)$, the hydrostatic pressure $P(t)$, the average coordination number $Z(t)$, and the volume fraction $\phi(t) = \mathrm{const}$ (maintained constant during the simple shear transformation). 
The pressure in the granular medium is estimated as the trace of the stress tensor $P=(\sigma_{xx} + \sigma_{yy} + \sigma_{zz})/3$, where $\sigma_{ii}=(\mathbf{F}_{+i}\cdot\mathbf{n}_{+i}+\mathbf{F}_{-i}\cdot\mathbf{n}_{-i})/2S_i$ with $S_i$ being the area of face $i=(x,y,z)$. It is computed from the total forces $\mathbf{F}_{\pm i}$ exerted by the particles on the faces defined by their normal vectors $\mathbf{n}_{\pm i}$.
We checked that this estimation of the confining pressure is consistent, by less than $0.1$\% relative difference, with that computed from the trace of the Love-Weber stress tensor~\citep{Makse2004,Thornton2000,Smilauer2021}.
The onset of jamming~\citep{OHern2003} (which corresponds to the onset of a pressure in the sample, at which it gets an elasticity) is reached at around a volume fraction $\phi_c = 0.595 \pm 0.003~(0.5\%)$ on average~\citep[p.111]{Faulconnier2023}, over the entire dataset presented in this study.
The shear stress $\tau$ is computed by summing all the reactions applied by the particles on the side walls, which are equivalent to two resisting moments computed at the center of rotation of the walls. 
The moments $\mathbf{M}_{+x}^{Oz}$ and $\mathbf{M}_{-x}^{Oz}$, exerted on the two opposite faces $\mathbf{n}_{\pm x}$, are used to compute the magnitude of an equivalent shear stress ${\tau=-2(\mathbf{M}_{+x}^{Oz}\cdot\mathbf{z}+\mathbf{M}_{-x}^{Oz}\cdot\mathbf{z})/V}$ exerted in-plane on the top wall $\mathbf{n}_{+z}$, where $\mathbf{z}$ is the Cartesian vector. Here, as a crude check, we verify that the work done by the shear force (the injected energy) equals the total energy per unit volume in the granular medium computed by the DEM, $\int{\tau d\gamma}=e$.


\section{Results}\label{sec:results}


\subsection{Evolution of shear stress}

Once the sample is set under pressure at a given packing fraction, a quasi-harmonic simple shear strain shown in Fig.~\ref{fig:1_sketchs_examples}a is imposed on it as described in Sec.~\ref{sec:method}.
The data in Fig.~\ref{fig:1_sketchs_examples}b present the evolution of the shear stress reaction between the walls of the container and the granular sample to accommodate such a motion for two examples at packing fraction $\phi=0.64$ and $\phi=0.70$. 
At such confinements, the average strain of the contacts~\citep{Roux2020} is $\langle h/a\rangle \approx 0.02$ and $\langle h/a\rangle \approx 0.04$, respectively, see Fig.~\ref{fig:A1_particle_deformation} in the Appendix Sec.~\ref{sec:A1_particle_deformation}.
In both cases, a nonlinear response is clearly evidenced in the stress-strain relation $\tau(\gamma)$, see Figs.~\ref{fig:1_sketchs_examples}d and \ref{fig:1_sketchs_examples}e. 
Indeed, we observe a hysteretic behavior in Fig.~\ref{fig:1_sketchs_examples}d at $\phi = 0.70$, with a noticeable dependence on the shear amplitude $\gamma_m$. For instance, following the evolution of the stress-strain curve $\tau(\gamma)$ with the largest amplitude, shown as the darkest blue curve in Fig.~\ref{fig:1_sketchs_examples}d, in the clockwise direction by starting from the left-bottom cusp point at $\gamma(t) = -0.15$, we can see that after a nonlinear growth, $\tau$ reaches a saturating point where it starts to increase linearly with respect to $\gamma$. 
Then $\tau$ reaches the right-top cusp point at $\gamma(t) = 0.15$ before pursuing the same track backward. 
The behavior at $(\phi,\gamma_m)=(0.70,0.15)$ must be contrasted to that at $(\phi,\gamma_m)=(0.64,0.15)$ presented in Fig.~\ref{fig:1_sketchs_examples}e, see the darkest red curve: the stress-strain curve follows a similar trend, except that the linear growth before reaching cusp points becomes even more nonlinear, with an additional stiffening inflection.
Such an inflection is less obvious in the case $(\phi,\gamma_m)=(0.64,0.10)$ seen in Fig.~\ref{fig:1_sketchs_examples}e, for which the hysteresis curves appear more similar to that at $(\phi,\gamma_m)=(0.70,0.15)$. 
Also in both cases seen in Figs.~\ref{fig:1_sketchs_examples}d and~\ref{fig:1_sketchs_examples}e, the hysteresis tends to shrink and transform into a linear elastic response at the smallest amplitudes.

All these observations thus highlight a symmetry of the roles played by the packing fraction, the confinement pressure and the shear strain amplitude on the mechanical response of the granular sample: large packing fractions, large confinement pressures and small strain amplitudes tend to linearize the stress-strain relation. This feature will be discussed further and exploited in more detail in the following sections.


\subsection{Evolution of pressure}\label{sec:Reynolds_pressure}

The evolution of the pressure during cyclic shear is plotted on Fig.~\ref{fig:1_sketchs_examples}c.
It shows that the pressure is fluctuating even though the cyclic shear transformation is isochoric.
This phenomenon has the same origin as the so-called Reynolds dilatancy, a usual fluctuation of the volume fraction $\phi$ when the shear transformation is performed at constant pressure, as for instance in~\citep{Xing2021}.
In the latter case, the granular medium tends to expand when subjected to shear strain.
In the case where the volume is constrained, the volume fraction is prescribed and the dilatancy is therefore restrained, resulting in a fluctuation of the pressure.
This variation, called {\em Reynolds pressure} in~\citep{Ren2013}, can be described as  
\begin{equation}
\widetilde{P}(t) = P + \delta P(t) = P_0 + R \gamma^2(t),
\label{eq:Reynolds_pressure}
\end{equation}
where $P$, $P_0$ and $\delta P$ correspond to the time average, the static value and the fluctuation of the pressure, respectively.
Because of the quadratic dependency in shear deformation, the static pressure $P_0$ (i.e. the pressure at zero strain, $\gamma_m=0$) and the mean pressure $P=P_0+\langle R\gamma^2\rangle$ are not equal; however, they remain close for small strain amplitudes, $R\gamma_m^2\ll P$.
The {\em Reynolds coefficient} $R$ was first introduced by Ren {\em et al} in~\citep{Ren2013}, when it was demonstrated to be related to the volume fraction, $R = A(\phi_c - \phi)^n$ with $n \simeq -3.3$ for disk packings in 2D and $A$ having the dimension of an elasticity.
It is worth pointing out that the behavior of $R$ was examined below jamming in~\citep{Ren2013}, $\phi<\phi_c$, while our study is conducted in a 3D packing of spherical grains above jamming, $\phi>\phi_c$, which is likely to provide a different scaling of $R$ with respect to the packing fraction.
In our case, we find that $R \propto (\phi-\phi_c)^n$ with $n \simeq 0.56$, see Fig.~\ref{fig:A2_nl}a in the Appendix Sec.~\ref{sec:A2_nl}.
In an alternative manner, a dependency of the pressure $P$ with the shear strain $\gamma$ was also captured by the constitutive model derived by Luding {\em et al} in~\citep{Krijgsman2013}, which demonstrates that the pressure fluctuations in a 2D granular material under cyclic pure shear result from structural anisotropy, coupling volumetric and deviatoric components in the stress tensor.

\begin{figure}[t]
\centering
\includegraphics[width=\linewidth]{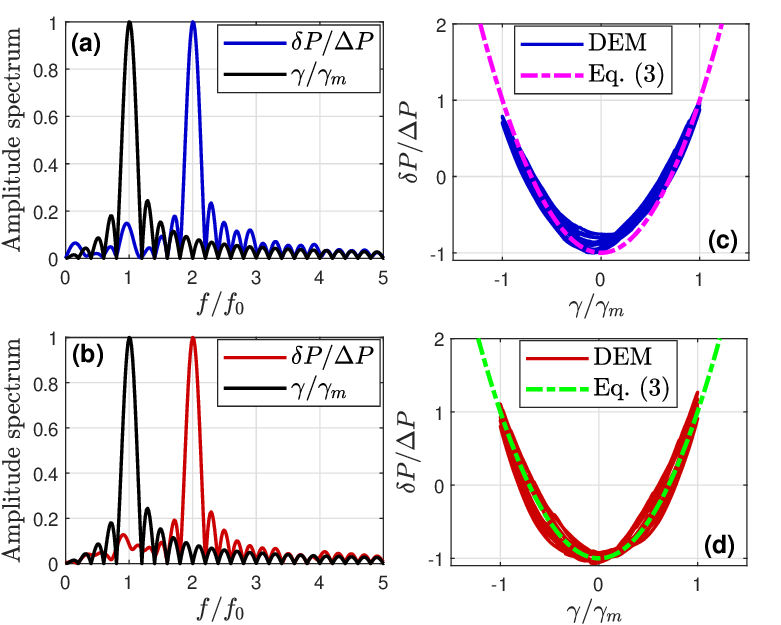}
\caption{
\textbf{(a,b)} Spectra of the shear strain and the pressure fluctuation and \textbf{(c,d)} pressure fluctuation as a function of the shear strain, for the two examples shown in Fig.~\ref{fig:1_sketchs_examples}: \textbf{(a,c)} $\phi = 0.70$ and \textbf{(b,d)} $\phi = 0.64$, both at at $\gamma_m=0.15$. Data in \textbf{(c,d)} are compared to a quadratic behavior, with \textbf{(c)} $R = 7.54$~GPa and \textbf{(d)} $R = 4.49$~GPa, where $\Delta P=R\gamma_m^2/2$ is the peak amplitude of the pressure fluctuations, see Eq.~\ref{eq:Reynolds_pressure}.}
\label{fig:2_Reynolds_pressure}
\end{figure}

In Fig.~\ref{fig:2_Reynolds_pressure}, we check the relevance of Eq.~\ref{eq:Reynolds_pressure} for the two examples shown in Fig.~\ref{fig:1_sketchs_examples}. 
To this end, we introduce the peak amplitude of the pressure fluctuations, $\Delta P=R\gamma_m^2/2$, stemming from the quadratic dependency $\gamma^2(t)=\gamma_m^2\sin^2{(\omega_0t)}=(\gamma_m^2/2)[1-\cos{(2\omega_0t)}]$.
First, Figs.~\ref{fig:2_Reynolds_pressure}a and \ref{fig:2_Reynolds_pressure}b reveal that the spectral component of the normalized pressure fluctuation $\delta P/\Delta P$ is mostly located at twice the driving frequency $f_0$ prescribed by the quasi-harmonic shear strain $\gamma/\gamma_m$. 
Such a frequency conversion is typical of a quadratic dependency, consistent with the reliable quadratic fit shown in Figs.~\ref{fig:2_Reynolds_pressure}c and \ref{fig:2_Reynolds_pressure}d.
The estimated values of $R$ from these fits, $R=4.49$~GPa for $(\phi,P)=(0.64,257\mathrm{~MPa})$ and $R=7.54$~GPa for $(\phi,P)=(0.70,929\mathrm{~MPa})$, indicate that $R$ is a growing function of $P$ and $\phi$. In greater detail, Appendix Sec.~\ref{sec:A2_nl} reveals a power law of the pressure, $R \propto P^m$ with $m \simeq 0.37$, see Fig.~\ref{fig:A2_nl}b, consistent with the aforementioned power law of the packing fraction, $R \propto (\phi-\phi_c)^n$ with $n \simeq 0.56$, see Fig.~\ref{fig:A2_nl}a.

Finally, we observe in Fig.~\ref{fig:1_sketchs_examples}c that the lower the packing fraction or the pressure, the higher $\delta P/P$ is.
This observation is compatible with the evolution of the nonlinear behavior evidenced in Figs.~\ref{fig:1_sketchs_examples}d and~\ref{fig:1_sketchs_examples}e: the pressure being a state variable, its fluctuation with the shear deformation will likely make the mechanical response nonlinear.
Indeed, the elasticity of granular media relies on Hertz contact theory, see above in Sec.~\ref{sec:method}, which depends on the confining pressure, see below in Sec.~\ref{sec:EMT}: a significant fluctuation of the pressure implies a nonlinearity of the elasticity.
In the following, we will thus restrict our description to cases at sufficiently small $\delta P/P$, i.e. at large packing fractions and pressures, to deal with linear elasticity; next, we will account for additional nonlinear contributions at smaller packing fractions and pressures.


\subsection{Elasto-hysteretic behavior}\label{sec:elast_hyst_behavior}

\begin{figure}[t]
\centering
\includegraphics[width=1\linewidth]{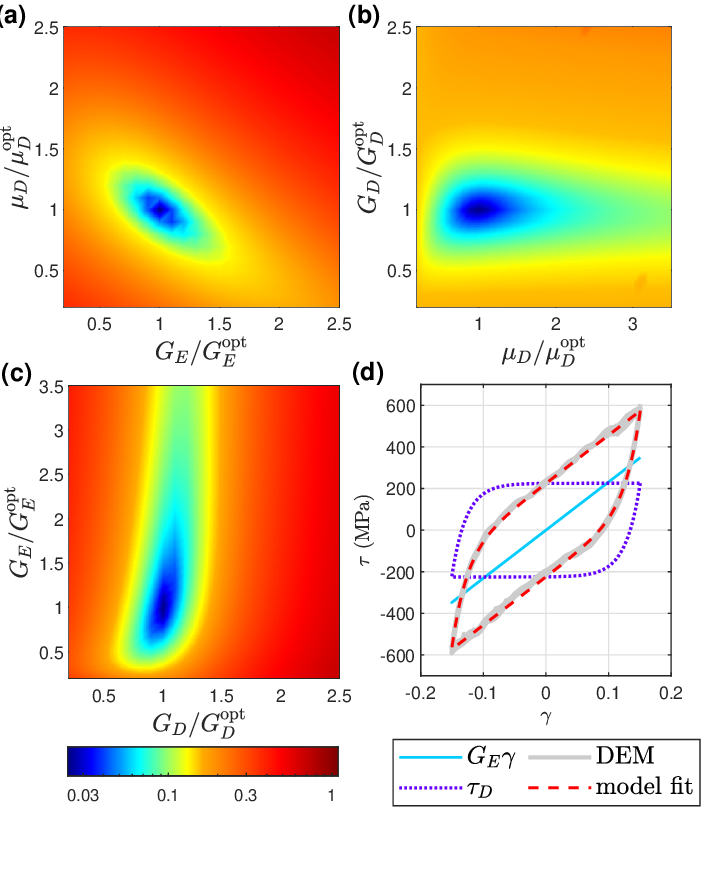}
\caption{\textbf{(a,b,c)} Maps of the RMSE between $\tau(\mu_D,G_D,G_E)$ and $\tau_\mathrm{DEM}$ plotted in the three planes $(\mu_D,G_E)$, $(\mu_D,G_D)$ and $(G_D,G_E)$, for the example $(\phi,\gamma_m)=(0.70,0.15)$ shown in Fig.~\ref{fig:1_sketchs_examples}d. The parameters are scaled by the optimal triplet $\mu_D^\mathrm{opt}=0.24$, $G_D^\mathrm{opt}=9.7$~GPa and $G_E^\mathrm{opt}=2.3$~GPa. \textbf{(d)} Stress-strain relation from DEM simulation superimposed on the fits. The purple dotted line represents the hysteretic Dahl contribution $\tau_D(\gamma)$, the solid blue line represents the linear contribution $G_E\gamma$, see Eq.~\ref{eq:model_reduit}, and the red dashed line is the sum of the two.}
\label{fig:3_parameters_identification}
\end{figure}

Here, we aim at reducing the complexity of the description of numerous particles subjected to shear, by identifying a low order approximation depending on essential parameters only.
To determine this model, we first focus on the example with a weak relative fluctuation of $P$, at $(\phi,\gamma_m)=(0.70,0.15)$ from Fig.~\ref{fig:1_sketchs_examples}d.
The stress-strain relation of this example is reproduced in Fig.~\ref{fig:3_parameters_identification}. At first glance, it can be seen that such a mechanical response is likely to be decomposed with a hysteretic response (a path-dependent saturating contribution) superimposed with a proportional response (a straight line with constant slope).
In the following, we describe the former with the Dahl model~\citep{Dahl1968,Dahl1977}, while the latter is expressed as a linear elastic term. 
Such an elasto-hysteretic decomposition was also used to describe the response of fibrous materials subjected to shear deformation~\citep{Piollet2016}.
The Dahl model is classically used in control engineering as an empirical model due to its facility of implementation; in particular, it was shown to constitute an alternative formulation of the Mindlin frictional interaction between a sphere and a plane~\citep{Peyret2017}.
The shear stress is thus expressed as
\begin{equation}
\tau(\gamma) = \tau_\mathrm{D}(\gamma,\dot{\gamma}) + G_E\gamma
\label{eq:model_reduit}
\end{equation}
where $G_E$ corresponds to a linear shear modulus and $\tau_D$ is the hysteretic Dahl stress. 
The latter has no explicit form, but it can be expressed in a compact form, as a first order time derivative nonlinear equation,
\begin{equation}
\dot{\tau}_D = G_D\dot{\gamma}\left[1 - \left(\frac{\tau_D}{\mu_D P}\right)\mbox{sign}(\dot{\gamma})\right]^{n_D},
\label{eq:model_dahl}
\end{equation}
where $G_D$, the first parameter of Dahl's model, is homogeneous with a shear modulus and corresponds to the slope of $\tau_D(\gamma)$ when $\tau_D = 0$. The second parameter, $\mu_D$, corresponds to a Coulomb friction coefficient so that $\mu_D P$ is the asymptotic value of $\tau_D$ at large shear strain.
The third parameter, exponent $n_D$, controls the shape of the curve, e.g., $n_D<1$ exhibits brittle behaviors, while $n_D>1$ renders ductile responses. In the rest of the study, we choose $n_D=1$ as the first attempt, in line with the literature~\citep{Piollet2016,Helmick2009,Majid2004}, to model elasto-plastic behaviors in between ductile and fragile.

The parameter identification consists in finding the optimal parameter triplet $(\mu_D^\mathrm{opt},G_D^\mathrm{opt},G_E^\mathrm{opt})$ that minimizes the difference between the reduced model defined in Eqs.~\ref{eq:model_reduit} and~\ref{eq:model_dahl}, and the steady mechanical response simulated with DEM.
More details on the process of identification is given in the Appendix Sec.~\ref{sec:A3_fit}.
To demonstrate the reliability of such identification, we show the root mean square error (RMSE) between $\tau(\mu_D,G_D,G_E)$ and $\tau_\mathrm{DEM}$ across three cross-sections of the parameters space, see Figs.~\ref{fig:3_parameters_identification}a, \ref{fig:3_parameters_identification}b, and \ref{fig:3_parameters_identification}c. 
It can be seen that a unique and well-defined global minimum exists in a wide region of parameters.
The qualitative agreement of the fit is shown in Fig.~\ref{fig:3_parameters_identification}d for the example $(\phi,\gamma_m) = (0.70,0.15)$, and the quantitative agreement is illustrated in more detail in the Appendix Sec.~\ref{sec:A3_fit} in Fig.~\ref{fig:A3_RMSE}.

\subsection{Energy dissipation}\label{sec:dissipation}

The hysteresis of the macroscopic stress-strain relation evidenced in Fig.~\ref{fig:3_parameters_identification} renders the mechanical response non-conservative because of a path-dependent mechanical loading/unloading. Likely the fingerprint of the microscopic inter-particle friction inside the granular sample, this mechanism therefore introduces energy dissipation. In this subsection we aim at defining and quantifying the damping properties of our system. In the following, the dissipated energy per volume unit and per cycle is measured as
\begin{equation}
E_\mathrm{diss}^\mathrm{cycle} = \frac{1}{n}\oint_{n~\mathrm{cycles}} \tau d\gamma,
\label{eq:E_diss}
\end{equation}
which represents the enclosed area inside the hysteresis loop. 
This quantity has to be contrasted with the maximum stored energy reached during a cycle.
The simplest estimation of that energy is~\citep{DelfosseRibay2004}
\begin{equation}
E_\mathrm{stored}^\mathrm{max} \simeq \frac{1}{2}\tau_{m}\gamma_{m}.
\label{eq:E_stored}
\end{equation}
The ratio between these two energies gives an estimation of the loss factor $\eta$ as
\begin{equation}
\eta = \frac{E_\mathrm{diss}^\mathrm{cycle}}{2\pi E_\mathrm{stored}^\mathrm{max}}. 
\label{eq:loss_factor}
\end{equation}

As an example, the definitions Eqs.~\ref{eq:E_diss}-\ref{eq:loss_factor} applied to the simulated strain-stress relation shown in Fig.~\ref{fig:3_parameters_identification}d indicate that the dissipated energy per cycle is $E_\mathrm{diss}^\mathrm{cycle} \simeq 114$ J/cm$^{3}$, whereas the maximum stored energy is $E_\mathrm{stored}^\mathrm{max} \simeq 45$ J/cm$^{3}$, leading to a loss factor $\eta \simeq 0.40$.
Such a level of damping is worthy of interest in engineering applications.
For the sake of comparison, an efficient viscoelastic material such as {\em Deltane 350} from {\em Paulstra}, a synthetic elastomer commonly used in the aeronautics field, ranges from a lower value $\eta\simeq0.16$ at $1$~Hz to a higher value $\eta\simeq0.88$ at $100$~Hz, reaching $\eta\simeq0.45$ at $10$~Hz~\citep[p.94]{Rouleau2013}. In contrast with the frequency-independent friction-based dissipation observed in our granular medium, the polymer material thus loses its efficiency at low frequency, in addition to being strongly temperature dependent~\citep[p.76]{Rouleau2013}.


\subsection{Parametric study}\label{sec:parametric}

We now look at generalizing our observations for various configurations in the parameter space. In particular, we seek to elucidate the possibility of an {\em optimal regime} for energy dissipation, in between low confinement pressure (where the particles freely flow on top of each other without rubbing) and high confinement pressure (where the particles firmly stick to form a sturdy elastic brick).
Three families of parameters are distinguished, namely (i) the particle properties, (ii) the state variables defining the packing, and (iii) the external parameters defining the driving excitation.

The {\em particle properties} include their Young’s modulus $E_p$, Poisson’s ratio $\nu_p$, mass density $\rho_p$, radius $r_p$, number $N_p$ and the inter-particle friction coefficient $\mu_p$.
For the sake of simplicity, fictional materials are considered, for which these parameters vary independently; however, they range within realistic values, like for the Young’s modulus $E_p$ reported in Tab.~\ref{tab:input_parameters}.
Without loss of generality, we assume that the Poisson’s ratio has a negligible effect (it only contributes to a second order correction of the elasticity, see Eqs.~\ref{eq:Hertz_Mindlin_n} and~\ref{eq:Hertz_Mindlin_t}) so that it is arbitrarily kept constant in this study, equal to its nominal value $\nu_p=0.27$. Also, the mechanical response should be independent of the mass of the particles at sufficiently low driving frequency (see below, in the quasi-static regime), the mass density considered in the following parametric study has a nominal value $\rho_p=7850$~kg/m$^{3}$.
In the same manner, it is likely that the size and the number of particles should not play a role at the macroscopic scale, in the long wavelength (low frequency, see below) continuum limit approximation. The insensitivity of the mass, the size and the number of the particles is probed at once successfully by simulating a sample made of particles with half the average radius (i.e. one eighth of the mass per particle) but the same sample's volume (i.e. eight times more particles) without a noticeable difference compared to the nominal case. This case is shown in Sec.~\ref{sec:EMT}.
Finally, the effect of inter-particle friction coefficient $\mu_p$ is probed separately, over more than a decade, in Sec.~\ref{sec:fric_eff}.

The {\em state variables} include the packing fraction $\phi$, the pressure $P$ and the coordination number $Z$, see Tab.~\ref{tab:input_parameters}.
These three quantities are related: according to the literature~\citep{OHern2003}, we checked (result not shown, see~\citep[p.111]{Faulconnier2023}) that $P \propto (\phi - \phi_c)^{3/2}$ and $(Z - Z_c) \propto (\phi - \phi_c)^{1/2}$ where $\phi_c = 0.595\pm0.003~(0.5\%)$ and $Z_c = 5.042\pm0.085~(1.7\%)$ are the values of the packing fraction and coordination number at the critical point where the jamming occurs.
Thus, imposing the packing fraction $\phi$ sets the average level of pressure $P$ and the average contact number $Z$.

Finally, the {\em external parameters} include the excitation angular frequency $\omega_0=2\pi f_0$ and shear amplitude $\gamma_m$, see Tab.~\ref{tab:input_parameters}.
The former is kept low enough to fulfill a quasi-static regime where the inertia of the grains does not make any contribution to the force balance. The latter is kept constant in the parametric study presented in Secs.~\ref{sec:EMT} and~\ref{sec:fric_eff} at a nominal value $\gamma_m = 0.15$ that ensures observing noticeable hysteresis in the stress-strain relation, as in Figs.~\ref{fig:1_sketchs_examples}d and~\ref{fig:1_sketchs_examples}e for instance. The dataset is then completed by probing different amplitudes, in Secs.~\ref{sec:dimless_analysis} and~\ref{sec:loss_factor}.

\begin{table}[t]
\caption{Parameter space.}
\centering
\begin{tabular}{c c c}
\hline
Parameter & value & unit\\
\hline
$E_p$ & $3 \sim 300$ & GPa\\
$\nu_p$ & $0.27$ & - \\
$\rho_p$ & $7850$ & kg/m$^{3}$ \\
$r_p$ & $(0.5 \sim 1.0) \pm 25\%$ & mm\\
$N_p$ & $1000 \sim 8000$ & - \\
$\mu_p$ & $0.017 \sim$ 0.6 & - \\
\hline
$\phi$ & $0.60 \sim 0.71$ & - \\
$P$ & $15 \sim 1300$ & MPa \\
$Z$ & $5.69 \sim 7.91$ & - \\
\hline
$\omega_0$ & quasi-static & rad/s \\
$\gamma_m$ & $0.005 \sim 0.15$ & - \\
\hline
\end{tabular}
\label{tab:input_parameters}
\end{table}

\begin{figure}[t]
\centering
\includegraphics[width=\linewidth]{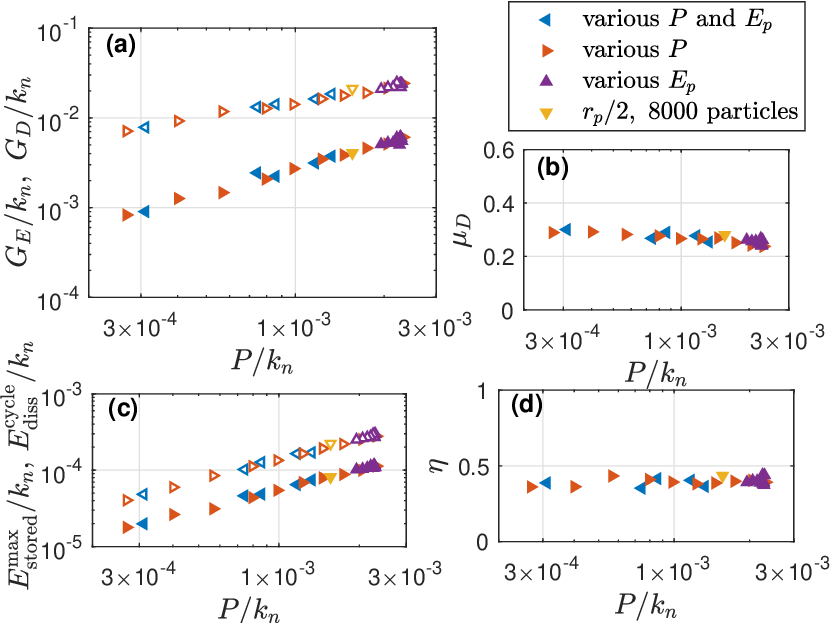}
\caption{
\textbf{(a)}: Dimensionless fitted parameters $G_D/k_n$ (open symbols) and $G_E/k_n$ (solid symbols) for various volume fractions $\phi$ and elasticities $E_p$ as functions of the dimensionless pressure $P/k_n$ at $\gamma_m=0.15$. 
\textbf{(b)}: Dahl friction coefficient versus dimensionless pressure for the same data set. 
\textbf{(c)}: Dissipated energy per cycle (open symbols) and maximum stored energy (solid symbols) computed from the hysteresis response using Eqs.~\ref{eq:E_diss} and \ref{eq:E_stored} respectively, as functions of the dimensionless pressure.
\textbf{(d)}: Loss factor computed from \ref{eq:loss_factor}, as a function of the dimensionless pressure. 
Legend symbols are common for \textbf{(a-d)}.
}
\label{fig:4_parametric_results}
\end{figure}

For each configuration of the parameter space, the system is simulated using DEM, from which we estimate the stress-strain curve in shear as described in Sec.~\ref{sec:method}. Then, an optimal triplet of parameters $(\mu_D,G_{D},G_{E})$ relying on the elasto-hysteretic reduced model are retrieved according to the protocol described in Sec.~\ref{sec:elast_hyst_behavior}.
These three parameters are plotted versus the state variable $P$ in Fig.~\ref{fig:4_parametric_results}a and~\ref{fig:4_parametric_results}b for a wide range of particle properties. For dimensional reasons, we normalize the pressure $P$ and the elasticity $G_{E,D}$ by the inter-particle contact stiffness (proportional to the Young's modulus of the particles’ material) in the normal direction, $k_n \propto E_p $, as defined in Eq.~\ref{eq:Hertz_Mindlin_n}.
Fig.~\ref{fig:4_parametric_results}a shows that both $G_E$ and $G_D$ increase with $P$, over a decade, resulting in two clearly defined but distinct slopes which will be analyzed in-depth in the Sec.~\ref{sec:discussion}.
In contrast, the effective Coulomb friction coefficient $\mu_D$ shown in Fig.~\ref{fig:4_parametric_results}b appears weakly dependent on the confinement pressure and the particle properties, with an average value $\langle \mu_D\rangle \simeq0.28$.

Finally, we look at the energetical performances of the granular packing, as defined in Eqs.~\ref{eq:E_diss}-\ref{eq:loss_factor}; again, we normalize the energies per unit volume by $k_n$ (i.e. by the particle's Young’s Modulus) for dimensional reasons.
Fig.~\ref{fig:4_parametric_results}c shows that both the dissipated and the maximum stored energy per cycle increase as functions of the confinement pressure, consistently with the elastic features of the effective model, $G_E$ and $G_D$. Also in agreement with the effective Coulomb friction coefficient $\mu_D$, their ratio (i.e. the loss factor), shown in Fig.~\ref{fig:4_parametric_results}d, appears weakly dependent on the pressure and the particle properties, situated close to the value $\eta=0.40$ already observed in Fig.~\ref{fig:3_parameters_identification}d.

\section{Discussion}\label{sec:discussion}


\subsection{Effective elasticity}\label{sec:EMT}

It is observed in Fig.~\ref{fig:4_parametric_results}a that the identified elastic moduli $G_D$ and $G_E$ are both correlated to the confinement pressure $P$. 
This behavior is expected in granular material.
Indeed, focusing on the low amplitude regimes, such that $\tau_D\ll\mu_DP$, the asymptote of the reduced model defined in Eq.~\ref{eq:model_reduit} and \ref{eq:model_dahl} indicates that ${\tau \rightarrow (G_E + G_D)\gamma}$. 
Hence, a total shear modulus of the granular packing, 
\begin{equation}
G_T = G_E + G_D,
\label{eq:shear_modulus_tot}
\end{equation}
appears relevant to describe the elastic response of the layer of particles in the infinitesimal limit.
It turns out that $G_T$ corresponds to the effective shear modulus defined by Walton's~\citep{Walton1987} effective medium theory (EMT) of granular materials.
This theory is based on Hertz contact mechanics~\citep{Johnson1985,Popov2010} under a mean field approximation (long wavelength continuum limit) of small perturbations and uniform strain (i.e. affine deformations~\cite{Makse2004}). 
For purely frictionless particles (tangential slip, $\mu_p=0$), the EMT's effective bulk and shear moduli are
\begin{eqnarray}
K_\mathrm{EMT}(P) & = & \frac{k_n}{12\pi}(\phi Z)^\frac{2}{3} \left(\frac{6\pi P}{k_n}\right)^\frac{1}{3}, \label{eq:EMT_K} \\
G_\mathrm{EMT}^0(P) & = & \frac{k_n}{20\pi}(\phi Z)^\frac{2}{3} \left(\frac{6\pi P}{k_n}\right)^\frac{1}{3}, \label{eq:EMT_Gzero}
\end{eqnarray}
respectively, where the contact stiffness $k_n$ between particles in the normal direction is defined in Eq.~\ref{eq:Hertz_Mindlin_n}, $Z$ is the coordination number, and $P$ is the mean (time average) hydrostatic pressure defined in Eq.~\ref{eq:Reynolds_pressure}.
For infinitely frictional particles (i.e. tangential stick, $\mu_p\rightarrow \infty$), the effective bulk modulus remains unchanged but the effective shear modulus,
\begin{equation}
G_\mathrm{EMT}^\infty(P) = \frac{k_n + (3/2)k_t}{20\pi}(\phi Z)^\frac{2}{3} \left(\frac{6\pi P}{k_n}\right)^\frac{1}{3},
\label{eq:EMT_Ginfty}
\end{equation}
is enhanced due to the contribution of the inter-particle contact stiffness in the tangential direction, $k_t$, defined in Eq.~\ref{eq:Hertz_Mindlin_t}. 
The expressions Eqs.~\ref{eq:EMT_K}-\ref{eq:EMT_Ginfty} indicate the typical scaling $G_\mathrm{EMT}^{0,\infty} \propto K_\mathrm{EMT} \propto E_p(P/E_p)^{1/3}$ passed on from Hertzian contact interaction.
Finally, it is worth reminding that the volume fraction and the coordination number are also pressure dependent~\citep{OHern2003}, $\phi(P)$ and $Z(P)$, as pointed out in Sec.~\ref{sec:parametric}.

\begin{figure}[t]
\centering
\includegraphics[width=\linewidth]{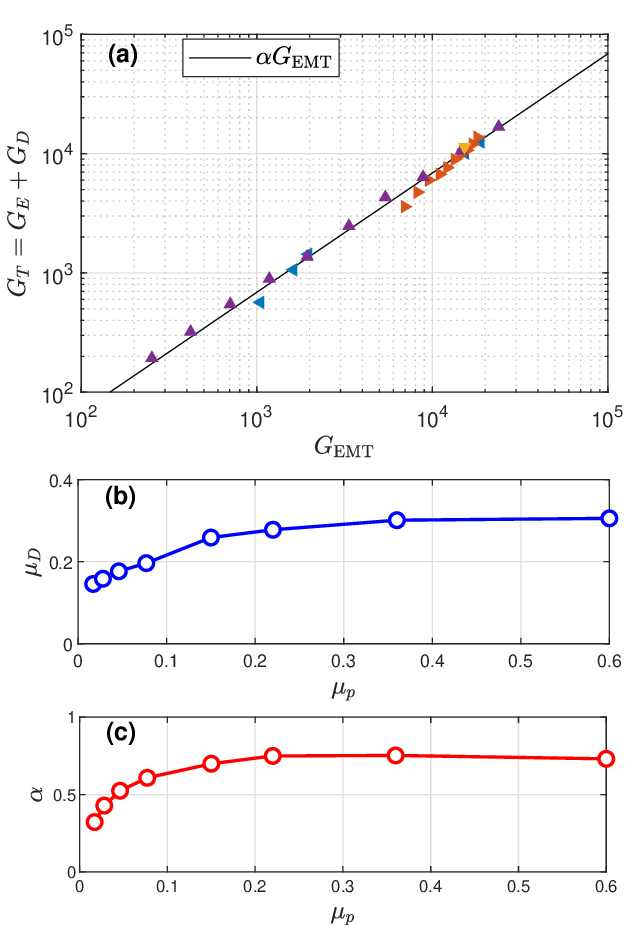}
\caption{
\textbf{(a)}: Total shear modulus $G_T=G_D+G_E$ versus effective shear modulus $G_\mathrm{EMT}^\infty$ (Eq.~\ref{eq:EMT_Ginfty}) with a linear fit of slope $\alpha = 0.687 \pm 0.073~(10.6\%)$. The same legend as in Fig.~\ref{fig:4_parametric_results}.
\textbf{(b,c)}: identified Dahl friction coefficient $\mu_D$ and coefficient $\alpha$ as functions of the inter-particle friction $\mu_p$. }
\label{fig:5_parametric_analysis}
\end{figure}

The plot in Fig.~\ref{fig:5_parametric_analysis}a of the total shear modulus versus the frictional effective modulus given in Eq.~\ref{eq:EMT_Ginfty} reveals a close correlation 
\begin{equation}
G_T = \alpha G_\mathrm{EMT}^\infty
\label{eq:GT_GEMT_relation}
\end{equation}
with a well-defined coefficient of proportionality, $\alpha=0.687 \pm 0.073~(10.6\%)$, on average over a dataset which is restricted to an inter-particle friction coefficient $\mu_p=0.15$.
In line with the EMT model, the value of the prefactor relies on the inter-particle friction coefficient: the formula given by Eqs.~\ref{eq:EMT_Gzero}-\ref{eq:EMT_Ginfty} indicates two asymptotic values $\alpha = G_\mathrm{EMT}^0/G_\mathrm{EMT}^\infty = 1/(1+3k_t/2k_n)\simeq 0.4$ for frictionless particles and $\alpha = G_\mathrm{EMT}^\infty/G_\mathrm{EMT}^\infty = 1$ for frictional ones.
The value of $\alpha$ at an intermediate $\mu_p = 0.15$ is thus consistent with these bounds.

\subsection{Effective friction}\label{sec:fric_eff}

We defined in Eq. \ref{eq:model_dahl} and reported in Fig.~\ref{fig:4_parametric_results}b a friction coefficient according to a Dahl model, as the ratio between the asymptotic stress at large strain and the confinement pressure.
In the field of the rheology of granular flow~\citep{gdrmidi2004}, a similar definition of an effective friction coefficient exists as the ratio of shear to normal force at the wall, or as the ratio of the shear stress to the pressure inside the material,
\begin{equation}
\mu(I) = \tau/P,
\label{eq:effective_fric}
\end{equation}
which is known to depend on a single dimensionless inertial number~\citep{Cruz2005,Jop2006}, $I=t_p/t_s$. 
The number $I$ is the ratio between two time scales, namely, the typical time $t_s = 1/\dot{\gamma}$ associated with shear rate, which correspond to the macroscopic deformation duration, and $t_p = r_p/\sqrt{P/\rho_p}$ that related to confinement, corresponding to the typical duration of microscopic rearrangement during which a grain moves by one neighbor size due the confining pressure~\citep{Jop2015}. 
The ratio $I$ describes the regime of a granular flow, from the quasi-static limit at low $I$ to the kinetic regime with multiple collisions at high $I$. 
A phenomenological equation~\citep{Jop2006} provides an estimation of the effective friction coefficient,
\begin{equation}
\mu(I) = \mu_s + \frac{\mu_g - \mu_s}{I_0/I + 1},
\label{eq:mu_de_I}
\end{equation}
where $I_0$ is a constant of the order of unity, which delimits the two asymptotic limits.
The effective friction coefficient value $\mu_g$ accounts for the existence of a gaseous phase at high inertia~\citep{Jop2015}, $I\gg I_0$.
In turn, the quasi-static value of the effective friction coefficient, $\mu_s$, mostly depends on the inter-particle friction coefficient $\mu_p$~\citep{Badetti2018}.
In practice, the estimated inertial number probed in our study remains small, $10^{-5} \leq I \leq 10^{-4}$, thanks to the low driving frequency (see Sec.~\ref{sec:method}), in such a way that the effective friction does not depend on the inertial number but only on the inter-particle friction coefficient, $\mu(I\ll I_0) \simeq \mu_s(\mu_p)$.
The analysis of the dependence of Dahl's effective friction coefficient, $\mu_D$ as a function of the inter-particle one, $\mu_p$, is shown in Fig.~\ref{fig:5_parametric_analysis}b, all the other parameters being fixed.
In particular, we observe that $\mu_D$ is an increasing function of $\mu_p$ with a saturation, in qualitative and quantitative agreement with the literature, see for instance Ref.~\citep{Badetti2018}.

Finally, we check the dependence of the prefactor $\alpha$ defined in Eq.~\ref{eq:GT_GEMT_relation} with respect to $\mu_p$.
As expected in Sec.~\ref{sec:EMT}, $\alpha$ should be a function of $\mu_p$, accounting for the transition between perfectly frictionless, $\alpha\simeq0.4$, and infinitely frictional particles, $\alpha=1$.
The observation reported in Fig.~\ref{fig:5_parametric_analysis}c is in qualitative agreement with this expectation, showing that the prefactor $\alpha$ increases with the inter-particle friction coefficient $\mu_p$ until saturating at a value of the order of unity.


\subsection{Dimensional analysis}\label{sec:dimless_analysis}

We determined in Sec.~\ref{sec:parametric} that the mechanical response depends on numerous particle features ($E_p$, $\nu_p$, $\rho_p$, $r_p$, $N_p$ and $\mu_p$), state variables ($P$, $Z$, $\phi$) and external parameters ($\omega_0$ and $\gamma_m$). However, Sec.~\ref{sec:EMT} and \ref{sec:fric_eff} show that most of these parameters are linked with each other. Here, we aim at reducing the dimension of the parameter space, by making use of the observations presented in Sec.~\ref{sec:elast_hyst_behavior}. In particular, we use the fact that the stress strain response is well described by an elasto-hysteretic model given by Eq.~\ref{eq:model_reduit} and \ref{eq:model_dahl}. 
We thus choose to rewrite these equations in dimensionless form using the following quantities:
\begin{gather}
\Gamma = \frac{\gamma}{\gamma_{m}},~\mathrm{T} = \frac{\tau}{G_T\gamma_{m}},\nonumber\\
\xi = \frac{\mu_D P}{G_T\gamma_m}~\mathrm{and}~\Psi = \frac{G_E}{G_T} =  1-\frac{G_D}{G_T},\label{eq:dimless_par}
\end{gather}
with $\Gamma$ and $\mathrm{T}$ the dimensionless shear strain and stress, respectively normalized by the amplitude of the shear strain and a typical shear stress proportional to the total shear modulus $G_T$ defined in Eq.~\ref{eq:shear_modulus_tot}.
In line with this choice, it stands to reason that Dahl's effective friction coefficient $\mu_D$ can be represented by a non-dimensional parameter $\xi$, the {\em first parameter} of our model, resulting from the competition between the typical shear stress imposed on the granular medium, $G_T\gamma_m$, and a Mohr-Coulomb-like threshold $\mu_D P$.
According to Fig.~\ref{fig:4_parametric_results}, the parameter $\xi$ is essentially a function of $P/k_n$ and $\gamma_m$.
In turn, the elastic term added to the hysteretic contribution, see Eqs.~\ref{eq:model_reduit} and \ref{eq:model_dahl}, can be described by a second parameter, $\Psi$, which stands as a {\em partition parameter} quantifying the ratio between the purely elastic shear modulus, $G_E$, and that stemming from the hysteretic response, $G_D$.
According to the normalizations defined in Eq.~\ref{eq:dimless_par}, the model Eqs.~\ref{eq:model_reduit} and \ref{eq:model_dahl} is rewritten as
\begin{equation}
\dot{\mathrm{T}}=\dot{\Gamma}\left[1-\left(\frac{1-\Psi}{\xi}\right)
(\mathrm{T}-\Psi\Gamma)\mathrm{sign}(\dot{\Gamma})\right],
\label{eq:dimless_eq}
\end{equation}
revealing that the mechanical response $\mathrm{T}(\Gamma)$ depends only on two parameters, $\xi$ and $\Psi$. 
From a physical point of view, the parameter $\xi$ provides information on the degree of hysteresis (related to frictional stress, $\mu_DP$) and elasticity (related to elastic stress, $G_T\gamma_m$) of the response, suggesting that $\Psi$ should also be a function of $\xi$.

\begin{figure}[t]
\centering
\includegraphics[width=\linewidth]{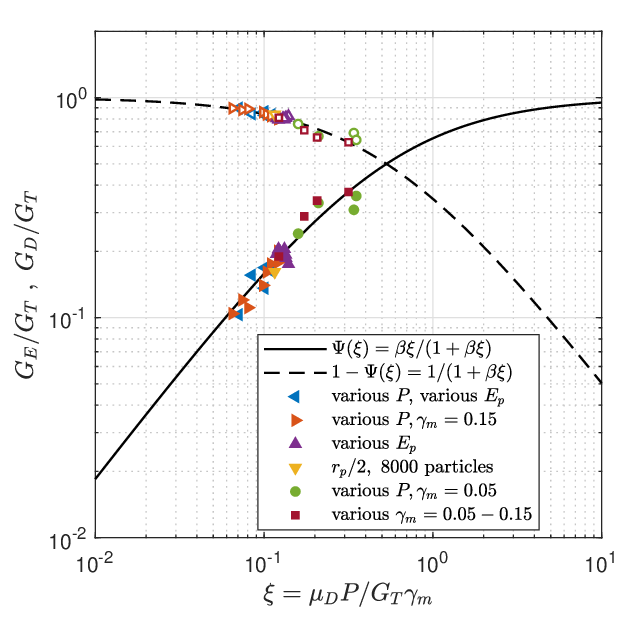}
\caption{Partition parameter $\Psi(\xi)$ as defined in Eq.~\ref{eq:dimless_par},
as a function of the dimensionless parameter $\xi = \mu_DP/G_T\gamma_m$. 
The solid line is the Ansatz introduced in Eq.~\ref{eq:Psi_xi} with $\beta=1.882 \pm 0.254~(13.5\%)$ best fitting the dataset.}
\label{fig:6_partition_function}
\end{figure}

Such an expectation is checked in Fig.~\ref{fig:6_partition_function}, which reveals a correlation between $\Psi$ and $\xi$.
In practice, the dependency is probed in the small-to-moderate values of $\xi$ only (see Fig.~\ref{fig:1_sketchs_examples}d for instance), where a hysteresis is clearly observable and leads to accurate estimations of the triplet $(\mu_D,G_D,G_E)$ i.e. of a pair $(\Psi,\xi)$. For larger $\xi$ (i.e. high pressure $P$ or low amplitude $\gamma_m$), the mechanical response tends to an elastic one, which prevents distinguishing the elasto-hysteretic contributions and leads to imprecise estimations of $\Psi$ and $\xi$.
However, considering that the granular medium becomes purely elastic in the high $\xi$ regime provides the information that the partition parameter $\Psi$ saturates at 1.
Hence, the simplest saturating function takes the form 
\begin{equation}
\Psi(\xi) = \frac{\beta\xi}{1+\beta\xi},
\label{eq:Psi_xi}
\end{equation}
where $\beta$ is an adjusting parameter. 
Fitting the Ansatz given in Eq.~\ref{eq:Psi_xi} to both $G_E$ and $G_D$ indicates that $\beta=1.882 \pm 0.254~(13.5\%)$. 
As intuited, the relation Eq.~\ref{eq:Psi_xi} thus shows that the model described by Eq.~\ref{eq:dimless_eq} depends only on the unique parameter $\xi$. 
Note that for the sake of completeness, the dataset shown in Fig.~\ref{fig:6_partition_function} is supplemented by simulations at various amplitudes, $0.05 \leq \gamma_m \leq 0.15$, in contrast with Fig.~\ref{fig:4_parametric_results} where all the simulations are done at $\gamma_m = 0.15$ only.


\subsection{Loss factor}\label{sec:loss_factor}

\begin{figure}[t]
\centering
\includegraphics[width=\linewidth]{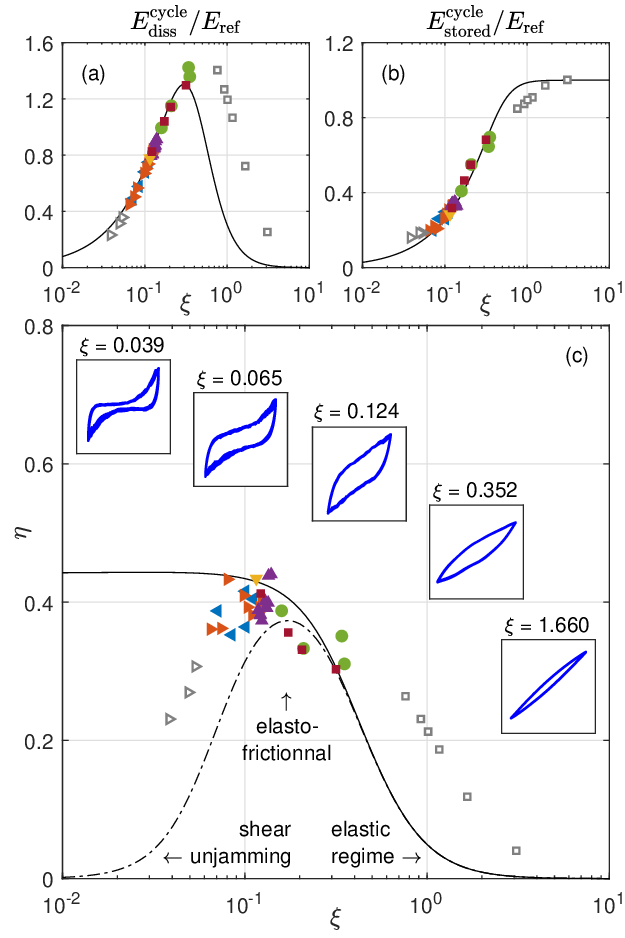}
\caption{
\textbf{(a)} Dissipated energy per cycle per volume unit, see Eq.~\ref{eq:E_diss}, as a function of $\xi$, normalized to the reference enregy defined in Eq.~\ref{eq:E_ref}. 
\textbf{(b)} normalized stored energy per cycle per volume unit, see Eq.~\ref{eq:E_stored}, versus $\xi$.
\textbf{(c)} Loss factor $\eta$ versus $\xi$, see Eq.~\ref{eq:loss_factor}. 
The legend of the colored markers is the same as in Fig.~\ref{fig:6_partition_function}.
Grey hollow symbols are supplementary points from DEM calculations at various small $\gamma_m$ (hollow squares) and various low pressures (hollow triangles). The solid line is the reduced model Eq.~\ref{eq:model_reduit}, while the dashed line stands for the improved reduced model Eq.~\ref{eq:model_reduit_NL}.  
}
\label{fig:7_energies_loss_factor}
\end{figure}

Here, we provide details about energy dissipation, in line with Sec.~\ref{sec:dissipation}, in the light of the dimensional analysis provided in Sec.~\ref{sec:dimless_analysis}.
In particular, we aim at probing how the dissipated energy Eq.~\ref{eq:E_diss}, the stored energy Eq.~\ref{eq:E_stored} and the loss factor Eq.~\ref{eq:loss_factor} depend on the non-dimensional parameter $\xi$.
To this end, we compare in Fig.~\ref{fig:7_energies_loss_factor} the data simulated by DEM to the features of the reduced model. The mechanical response of the model, $\mathrm{T}(\Gamma)$, is determined via the explicit numerical time integration of Eqs.~\ref{eq:dimless_eq}-\ref{eq:Psi_xi} for every $\xi$. Afterwards, the energetic features are extracted according to Eqs.~\ref{eq:E_diss}-\ref{eq:loss_factor}.
Note that for dimensional purposes, energies will be represented with respect to a reference energy,
\begin{equation}
E_\mathrm{ref} = \frac{1}{2}G_T\gamma_{m}^2,
\label{eq:E_ref}
\end{equation}
which stands for the energy stored by a purely elastic medium, $\tau = G_T\gamma$.

As expected from the analysis of Sec.~\ref{sec:dimless_analysis}, we observe in Fig.~\ref{fig:7_energies_loss_factor} that both energies and the loss factor are well correlated to $\xi$. The agreement between the data simulated by DEM and the reduced model is both qualitative and quantitative in the moderate range of $\xi$.
From low-to-moderate $\xi$, the non-dimensional dissipated energy rises because increasing the pressure, for instance, raises the frictional dissipative force $\mu_D P$; concurrently, the medium transits from a loose (at low Coulomb threshold) to a dense (at high Coulomb threshold) state, thus progressively gaining elasticity and the ability to store energy.
In contrast, at higher pressure the particles stick firmly to each other and the sample tends to store energy as a purely elastic medium: from moderate-to-high $\xi$, the dissipated energy vanishes while the non-dimensional stored energy tends to unity.
Consistently with these trends, the loss factor $\eta$ appears relatively constant in the moderate range of $\xi$ (see the colored markers in Fig.~\ref{fig:7_energies_loss_factor}c, $0.06\leq \xi \leq 0.4$), in agreement with the observation in Fig.~\ref{fig:4_parametric_results}d that $\eta$ appears fairly independent of the pressure within the same range of parameters.
The trends also reveal a decrease of $\eta$ which is characteristic of an elastic medium at high $\xi$, but in a range that is not accessed by the simulations. However, the reduced model reveals a plateau at low $\xi$, but again not evidenced by the DEM simulations.

On the {\em one hand}, we supplement six data at high $\xi$ (white squares in Fig.~\ref{fig:7_energies_loss_factor}) which tend to confirm the decay predicted by the model, namely, a jammed granular medium under high pressure behaves as an elastic solid (weak dissipation, high stored energy, and thus low loss factor).
On the {\em other hand}, the trend of the three data added at low $\xi$ (white triangles in Fig.~\ref{fig:7_energies_loss_factor}) reveals a decay of $\eta$ which clearly differs from the prediction given by the reduced model. 
Interestingly, such a decay is concomitant with the nonlinear stiffening inflection observed near the cusp points in Fig.~\ref{fig:1_sketchs_examples}e, at low $P$ and high $\gamma_m$, concurrently with the significant relative fluctuation of the pressure shown  in Fig.~\ref{fig:1_sketchs_examples}c.
The buildup of such additional nonlinearity in the stress-strain response at low $\xi$ is clearly evidenced in the examples $\xi = 0.124$ to $\xi = 0.039$ given in the inset of Fig.~\ref{fig:7_energies_loss_factor}c.


\subsection{Weakly nonlinear response}\label{sec:NL_response}

As expected in the conclusions of Sec.~\ref{sec:Reynolds_pressure} and from the developments of Sec.~\ref{sec:EMT}, the fluctuation of the pressure due to the restrained dilatancy described in Eq.~\ref{eq:Reynolds_pressure} is a potential source of nonlinearity of the elasticity, and thus an opportunity to improve the reduced model defined in Eqs.~\ref{eq:model_reduit} and \ref{eq:model_dahl}.
In the first qualitative attempt, aimed at capturing such a nonlinearity in the stress-strain relation, we will only consider a correction in the effective elasticity defined in Eqs.~\ref{eq:EMT_K}-\ref{eq:EMT_Ginfty}, by arbitrarily leaving unchanged the pressure in the Coulomb stress, $\mu_DP=\mathrm{const}$ see Eq.~\ref{eq:model_dahl}.
The first order Taylor expansion of the  shear modulus defined in Eq.~\ref{eq:GT_GEMT_relation} provides 
\begin{eqnarray}
G_T(\widetilde{P}) &=& G_T(P)\times \left( 1 +  \delta P/P \right)^{1/3},\nonumber\\
&\simeq& G_T(P)\times \left( 1 + \rho \Gamma^2 \right),\nonumber\\
\rho &=& R\gamma_m^2/3P = 2\Delta P/3P.\label{eq:nonlinear_elasticity}
\end{eqnarray}
The parameter $\rho$ is proportional to the relative pressure fluctuation and quantifies the nonlinearity of the response, see the comments of Fig.~\ref{fig:1_sketchs_examples}e.
At the lowest order, the expansion Eq.~\ref{eq:nonlinear_elasticity} thus leads to a nonlinear cubic correction term in the reduced model Eq.~\ref{eq:model_reduit},
\begin{eqnarray}
\tau(\gamma) &\simeq& \tau_\mathrm{D}(\gamma,\dot{\gamma}) + G_E\gamma + G_{NL}\gamma^3,\nonumber\\
G_{NL} &=& \rho G_T/\gamma_m^2 = RG_T/3P.\label{eq:model_reduit_NL}
\end{eqnarray}
It is worth mentioning that this additional term is conservative, such that it only enhances the stored energy and does not affect the dissipated energy; it thus likely lowers the loss factor predicted by the reduced model, as observed in Fig.~\ref{fig:7_energies_loss_factor}c at low $\xi$ (low pressures $P$ or high amplitudes $\gamma_m$).

\begin{figure}[t]
\centering
\includegraphics[width=\linewidth]{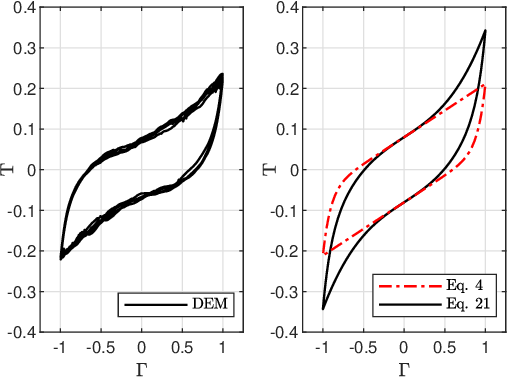}
\caption{
\textbf{(a)} Non-dimensional stress-strain curve of DEM simulation in the weakly nonlinear regime ($\phi=0.64$ at $\gamma_m=0.15$, such that $\xi = 0.080$ and $\rho =2\Delta P/3P\simeq 0.131$): the DEM simulated loss factor is $\eta = 0.368$.
\textbf{(b)} Prediction given by the reduced model Eq.~\ref{eq:model_reduit} (red dashed-dotted line, loss factor $\eta = 0.438$) and by the improved reduced model Eq.~\ref{eq:model_reduit_NL} (black solid line, loss factor $\eta = 0.270$). Compared to the DEM simulation, the former does not show the characteristic inflection near the cusp points, underestimates the peak amplitude of the stress by $11$\% and overestimates the loss factor by $19$\%. The latter shows this inflection qualitatively, but overestimates the peak amplitude of the stress by $45$\% and underestimates the loss factor by $27$\%.
}
\label{fig:8_nonlinear_response}
\end{figure}

As an example, we examine the contribution of such a correction for the case $(\phi,\gamma_m)=(0.64,0.15)$ shown in Fig.~\ref{fig:1_sketchs_examples}e, for which $\xi = 0.080$ and $\rho = 2\Delta P/3P\simeq 0.131$, and whose DEM simulation is reproduced in Fig.~\ref{fig:8_nonlinear_response}a.
As seen in Fig.~\ref{fig:8_nonlinear_response}b, the stress-strain relation predicted by the model defined in Eq.~\ref{eq:model_reduit}, see the red dashed-dotted curve, suffers from poor agreement: the identification of the parameters described in Sec.~\ref{sec:elast_hyst_behavior} inherently matches the dissipated energy (see the area within the curve) and acquires overall elasticity well, but fails to reflect either the slope at small strains (near $\Gamma=0$) or the characteristic inflections at large strains (near $\Gamma=1$).
The Eq.~\ref{eq:model_reduit_NL} provides a more convincing qualitative agreement, with $G_{NL} = \rho G_T/\gamma_m^2$ estimated by using the value of $G_T$ fitted by Eq.~\ref{eq:model_reduit} and the value $\rho=0.131$ estimated from the pressure fluctuation, see the solid black line in Fig.~\ref{fig:8_nonlinear_response}b. However, the quantitative agreement is less obvious: the original model Eq.~\ref{eq:model_reduit} overestimates the loss factor simulated by DEM by $19$\% whereas the improved model Eq.~\ref{eq:model_reduit_NL} underestimates it by $27$\%.

A better quantitative agreement can be reached by adjusting the parameters of Eq.~\ref{eq:model_reduit_NL} independently, since for instance the value of $G_T$ obtained with Eq.~\ref{eq:model_reduit} is not very relevant in the weakly nonlinear regime. However, the identification of a set of four parameters, $(\mu_D,G_D,G_E,G_{NL})$, introduces numerical uncertainties that require a protocol as careful as that detailed in Fig.~\ref{fig:3_parameters_identification}. An even more reliable improvement is to also consider the pressure fluctuation not only on the elastic contribution, as in Eq.~\ref{eq:model_reduit_NL}, but in the hysteretic part too. However, both these additional improvements are beyond the scope of our qualitative attempt.

Nevertheless, the dimensional analysis of Eq.~\ref{eq:model_reduit_NL}, derived as in Sec.~\ref{sec:dimless_analysis}, shows straightforwardly that it depends on the two non-dimensional parameters $\xi$ and $\rho$ only. Attempting to reduce further the description of the model, we consider that $\xi$ is essentially a function of $P/k_n$ and $\gamma_m$ and we probe the correlations of $\rho$ with the same two parameters, as detailed in Figs.~\ref{fig:A2_nl}c and~\ref{fig:A2_nl}d of the Appendix Sec.~\ref{sec:A2_nl}. Understanding such dependencies is out of the scope of the present paper. However, they allow estimating an improved version of the loss factor in a smaller parameter space, $\eta(\xi(P/k_n,\gamma_m),\rho(P/k_n,\gamma_m))$. This improvement is exemplified in Fig.~\ref{fig:7_energies_loss_factor}c (dashed-dotted black curve) for the nominal case $\gamma_m=0.15$ and the pressure dependent $\rho$ as defined in Fig.~\ref{fig:A2_nl}c. The plot shows a trend of the loss factor at low $\xi$ that agrees with DEM simulations qualitatively (see the hollow triangles calculated at low pressures), thus demonstrating that the lowering of $\eta$ likely results from the nonlinearity of the elasticity near the unjamming condition.

From the practical point of view, the decay of the loss factor $\eta$ at both low and high $\xi$ leaves a decade wide region, around $0.05<\xi<0.5$, where the loss factor is maximal and ranges between 20\% and 40\%. Technically, this provides an opportunity to tune the dissipation in order to target specific operating points, by adjusting for instance the confinement pressure to mitigate a given range of vibrations amplitudes efficiently. Interestingly, the optimal value $\eta\simeq40$\% is independent of the particle properties.


\section{Conclusion}\label{sec:conclusion}

In this paper, an elasto-hysteretic model describing a simply sheared granular medium confined in a box with rigid but moving walls has been studied using the discrete element method. In particular, it was shown that a single degree-of-freedom Dahl model coupled with an elastic response reliably describes the behavior simulated by DEM, in terms of the stress-strain relation, the stored and the dissipated energies, and the loss factor.
The well-known effective medium theory predicting the elasticity of granular media, and a macroscopic effective friction coefficient, were identified as the main parameters of this model, via an extensive parametric study derived as a function of the confining pressure and the features of the particles and the mechanical excitation.

Our study reveals that the elasto-hysteretic mechanical response of the granular medium mainly depends on a single dimensionless parameter, namely $\xi$, akin to a normalized Coulomb threshold.
This parameter allows quantifying {\em a priori} the extent to which the mechanical response of our apparatus is hysteretic compared to elastic. The regime at a high value of $\xi$ (i.e. high confinement pressure or low shear amplitudes) is asymptotically elastic, such that the loss factor vanishes in this regime.
Nevertheless, the proposed model fails at a low value of $\xi$ (i.e. low pressure or high shear amplitudes) due to large pressure fluctuation resulting from the anisotropy of the contact network. This effect leads to an additional nonlinearity of the elasticity, which ultimately results in the decrease of the damping ratio in this regime.
Hence, in an intermediate range of $\xi$, the dissipation is dominated by an effective friction coefficient that is mostly independent of the particle features, whereas the stored energy is dominated by an effective elasticity that mostly depends on the confinement pressure and the elasticity of the particle. Thus the result is that the loss factor can be tuned by the confinement pressure, as its maximum is independent of the particles and operating point.

From the practical point of view, the reduced model derived in this study can be used in more complex configurations, such as a granular medium clamped in between elastically bending beams to form a composite structure intended to dissipate mechanical vibrations. Here, the advantage of a reduced model is to lower the complexity and the numerical cost of simulations coupling the discrete element method with the finite element method, hence allowing subsequent optimization processes, for instance. We are now considering designing a granular composite beam~\citep{Faulconnier2023}. Also, the seminal features of this study have been patented~\citep{Fichant2024} by some of our team in a separate work conducted in collaboration with {\em Safran Spacecraft Propulsion}, a major stakeholder of the Aeronautic Industry. The patented equipment is aimed at mitigating the vibrations of a cathode in the harsh environment of a spacecraft plasma engine, by using non-cohesive ceramic particles confined inside structural elements, as described in the present study.


\begin{appendix}

\renewcommand\thesection{A\arabic{section}}
\setcounter{section}{0}

\renewcommand\thesubsection{A\arabic{subsection}}
\setcounter{subsection}{0}

\renewcommand\theequation{A\arabic{equation}}
\setcounter{equation}{0}

\renewcommand\thefigure{A\arabic{figure}}
\setcounter{figure}{0}

\renewcommand\thetable{A\arabic{table}}
\setcounter{table}{0}

\section*{Appendix}


\subsection{Particle deformation}\label{sec:A1_particle_deformation}

\begin{figure}[b]
\centering
\includegraphics[width=\linewidth]{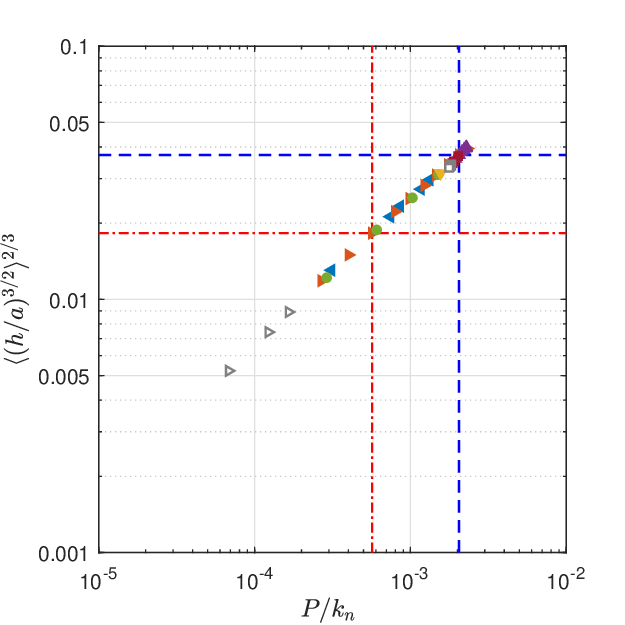}
\caption{Average contact strain as a function of the dimensionless pressure. Dashed-dotted red line and dashed blue line correspond to the computations from Figs.~\ref{fig:1_sketchs_examples}d and \ref{fig:1_sketchs_examples}e at $(\phi,\gamma_m) = (0.64,0.15)$ and $(\phi,\gamma_m) = (0.70,0.15)$, respectively. Legend  of the symbols is the same as in Fig.~\ref{fig:6_partition_function}.}
\label{fig:A1_particle_deformation}
\end{figure}

Provided that the contact between particles follows Hertz theory, the average contact strain in a jammed granular packing can be calculated according to Roux {\em et al.}~\citep{Roux2020}
\begin{equation}
\langle \left( h/a\right)^{2/3} \rangle^{3/2} = \left( \frac{3 \pi}{Z\phi}\right)^{2/3}\left( \frac{P}{k_n/2}\right)^{2/3},
\end{equation}
where $h$ corresponds to the contact overlap and $a$ to the particle diameter. In our study, the average contact strain 
ranges between $0.5\%$ and $4\%$, see Fig.~\ref{fig:A1_particle_deformation}.
It is worth mentioning that the high confinement pressure considered for the nominal configuration, $P/k_n\simeq0.2\%$, is chosen to accommodate the large shear deformation, $\gamma_m=0.15$, required to generate a clear hysteretic simulated response as in Fig.~\ref{fig:1_sketchs_examples}. It turns out that the typical normal stress within a contact is $(\sigma/k_n)\propto (P/k_n)^{1/3}\sim0.1$ according to the Hertz potential~\cite{Johnson1985}; very few materials can withstand such stress while remaining elastic, as for instance certain plastic materials and elastomers. However, the conclusions of the present study state that low shear amplitudes require low confinement pressure owing to the dependence of the mechanical response on the parameter $\xi\propto(P/k_n)^{2/3}\gamma_m^{-1}$, see Eq.~\ref{eq:dimless_par}. For instance, glass particles were safely used in a patented prototype~\cite{Fichant2024} designed to operate efficiently at moderate confinement pressures $P/k_n\ll1$ in the range of vibration amplitudes $\gamma_m\ll1$.


\subsection{Nonlinear parameter}\label{sec:A2_nl}

\begin{figure}[t]
\centering
\includegraphics[width=\linewidth]{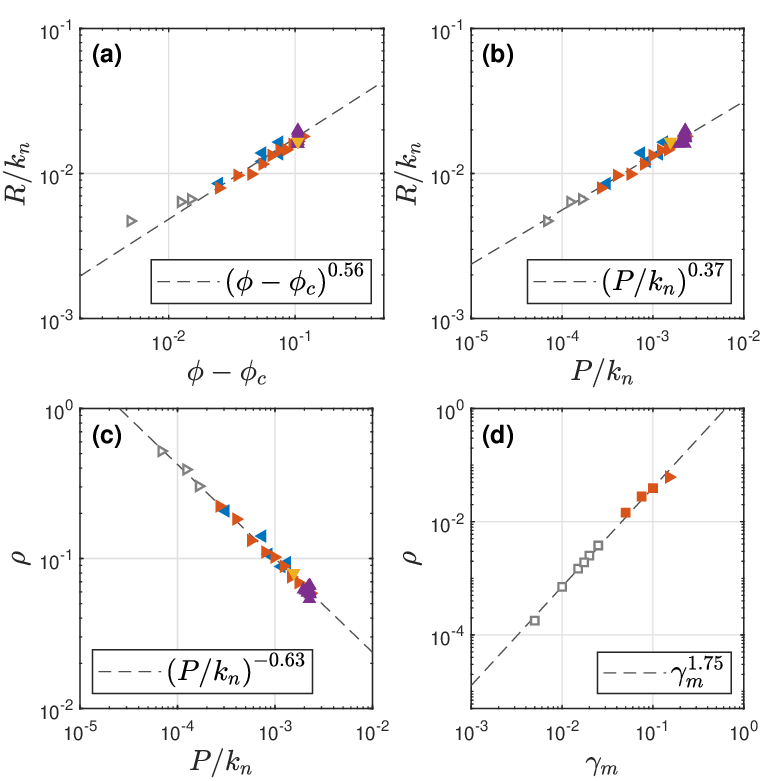}
\caption{\textbf{(a-b)} Scaling of the normalized Reynolds coefficient $R$ with respect
\textbf{(a)} to the packing fraction and
\textbf{(b)} to the dimensionless confining pressure, both at constant $\gamma_m = 0.15$.
\textbf{(c-d)} Scaling of the nonlinear parameter $\rho=2\Delta P/3P$ with respect
\textbf{(c)} to the dimensionless pressure at constant $\gamma_m = 0.15$ and
\textbf{(d)} to the shear strain amplitude at constant $k_n$ and $P$. 
Equations of the fits in dashed lines are $R/k_n = 0.06\left(\phi-\phi_c\right)^{0.56}$, $R/k_n = 0.18\left(P/k_n\right)^{0.37}$, $\rho = 1.33\times 10^{-3}\left(P/k_n\right)^{-0.63}$ and  $\rho = 2.22\gamma_m^{1.75}$, respectively.}
\label{fig:A2_nl}
\end{figure}

The effect of the nonlinearity resulting from the pressure fluctuation at low $\xi$ can be quantified by looking at the dependence of the Reynolds coefficient $R$ defined in Eq.~\ref{eq:Reynolds_pressure} with the state variables $P$ or $\phi$ (both are related), and at the dependence of the nonlinear parameter $\rho = 2\Delta P/3P$ defined in Eq.~\ref{eq:nonlinear_elasticity} with the pressure $P$ and the amplitude $\gamma_m$.
The results are displayed in Fig.~\ref{fig:A2_nl}, showing well defined trends: 
(i) $R$ is an increasing function of a power law of the relative packing fraction $(\phi-\phi_c)$;
(ii) also, $R$ is an increasing function of a power law of the confining pressure $P$;
(iii) $\rho$ is a decreasing function of $P$, meaning that loose packing leads to strong nonlinearity at a given amplitude $\gamma_m$;
(iv) $\rho$ is an increasing function of $\gamma_m$, meaning that nonlinearity increases classically with the amplitude of perturbation.


\subsection{Parameter fitting procedure}\label{sec:A3_fit}

\begin{figure}[b]
\centering
\includegraphics[width=\linewidth]{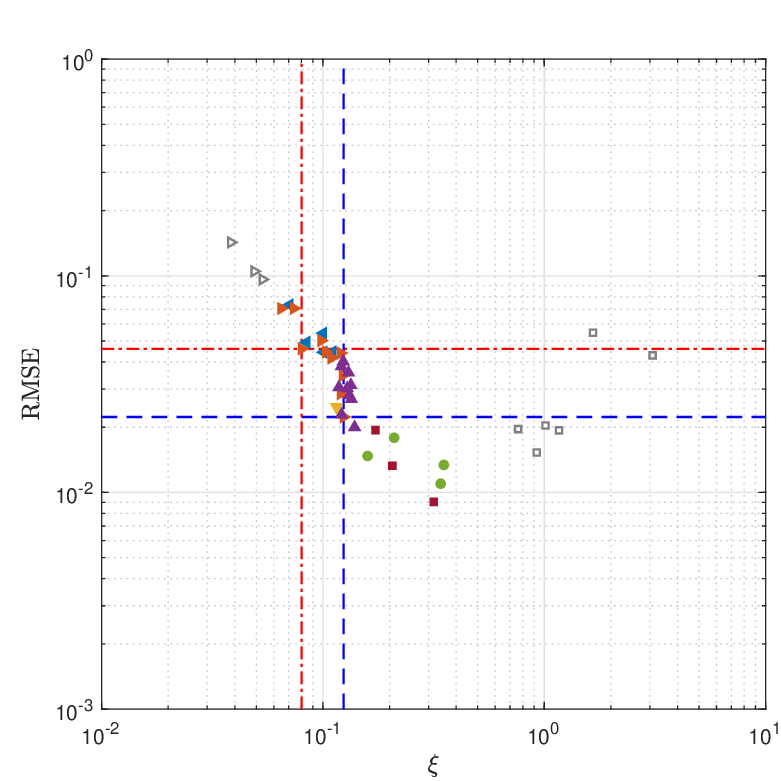}
\caption{Optimal RMSE between stress-strain relation given by the elasto-hysteretic reduced model Eq.~\ref{eq:model_reduit} and the simulations with DEM, as a function of the dimensionless parameter $\xi$.
The red dashed-dotted line and the dashed blue line correspond to the computations from Figs.~\ref{fig:1_sketchs_examples}d and \ref{fig:1_sketchs_examples}e at $\gamma_m = 0.15$, at $\phi = 0.64$ and $\phi = 0.70$, respectively. The legend of the symbols is the same as in Fig.~\ref{fig:6_partition_function}.}
\label{fig:A3_RMSE}
\end{figure}

The fitting of the parameters $(\mu_D,G_D,G_E)$ described in Secs.~\ref{sec:elast_hyst_behavior} and~\ref{sec:parametric} is performed in two steps: the determination of an initial guess $(\mu_D^\mathrm{ini},G_D^\mathrm{ini},G_E^\mathrm{ini})$ by successive steps, followed by one refinement to obtain an optimal triplet $(\mu_D^\mathrm{opt},G_D^\mathrm{opt},G_E^\mathrm{opt})$.
We first look for an initial ensemble $G_D^\mathrm{ini}(\mu_D)$ as a function of a finite number of friction coefficients taken in a finite interval $0.01\leq\mu_D\leq10$, which provides a hysteresis loop given by the model Eq.~\ref{eq:model_reduit} having the same dissipated energy as in the DEM simulation.
This step is made possible because, among the three parameters to be fitted, the dissipated energy does not depend on the conservative contribution associated with $G_E$.
In practice, we determine $G_D^\mathrm{ini}$ at a given $\mu_D$ by minimizing (using {\em Matlab}'s simplex algorithm {\em fminsearch}) the root-mean-square error (RMSE) between the area within one stress-strain close loop estimated (using {\em Matlab}'s numerical integration {\em trapz}) by the model Eq.~\ref{eq:model_reduit} and by the DEM simulation. This step provides a discrete numerical function, $G_D^\mathrm{ini}(\mu_D)$, with which we then search for a value $G_E^\mathrm{ini}(\mu_D)$ that minimizes the RMSE between the instantaneous stress response given by the model Eq.~\ref{eq:model_reduit} and that of the simulation, at any given $\mu_D$. Finally, the initial guess $\mu_D^\mathrm{ini}$ is the value that provides the lowest aforementioned RMSE.
The triplet ($\mu_D^\mathrm{ini},G_D^\mathrm{ini},G_E^\mathrm{ini}$) thus constitutes a reliable starting point to perform a 3-parameter minimization of the RMSE between the model Eq.~\ref{eq:model_reduit} and the simulated instantaneous stress response, defined by the following cost function: 
\begin{equation}
C(\mu_D,G_D,G_E) = \frac{\sqrt{\langle\left[\tau_\mathrm{DEM}(t) - \tau_\mathrm{model}(t)\right]^2\rangle}}{\sqrt{\langle\tau_\mathrm{DEM}(t)^2\rangle}}.
\label{eq:A3_RMSE}
\end{equation} 
This last step leads to a refined estimation of the optimal triplet $(\mu_D^\mathrm{opt},G_D^\mathrm{opt},G_E^\mathrm{opt})$.
We provide in Fig.~\ref{fig:A3_RMSE} the final value of the RMSE, defined in Eq.~\ref{eq:A3_RMSE} and estimated at the optimal triplet, as a function of the dimensionless parameter $\xi$ for all the data presented in this study. The model Eq.~\ref{eq:model_reduit} fits reliably in the range of $\xi\geq0.1$ where the {\em linear} elasto-hysteretic regime prevails: the RMSE remains below $5$\%. However, the fits provides poorer results at low $\xi$ where the {\em nonlinear} elastic contribution detailed in Sec.~\ref{sec:NL_response} becomes noticeable; in particular, the RMSE is larger than $10$\% for the three data introduced in Sec.~\ref{sec:NL_response} (see the white triangles at the lowest $\xi$), demonstrating the need to improve the model Eq.~\ref{eq:model_reduit} as in Eq.~\ref{eq:model_reduit_NL}.


\subsection{Size and number of particles}\label{sec:A4_check_part_num_size}

\begin{table}[b]
\caption{Parameters values for various combinations of particles' number, size and polydispersity; the case (b) is the nominal configuration.}
\centering
\begin{tabular}{|l|c|c|c|c|}
\hline
case & $N_p$ & $r_p$ (mm) & $dr_p/r_p$ ($\%$) \\ 
\hline
(a) & 125 $(5 \times 5 \times 5)$ & 2 & 25 \\
\hline
(b) & 1000 $(10 \times 10 \times 10)$ & 1 & 25 \\ 
\hline
(c) &  8000 $(20 \times 20 \times 20)$ & 0.5 & 25 \\ 
\hline
(d) & 1000 $(10 \times 10 \times 10)$ & 1 & 40 \\ 
\hline
\end{tabular}
\label{tab:A4_check_part_num_size}
\end{table}

\begin{figure}[t]
    \centering
    \includegraphics[width=\linewidth]{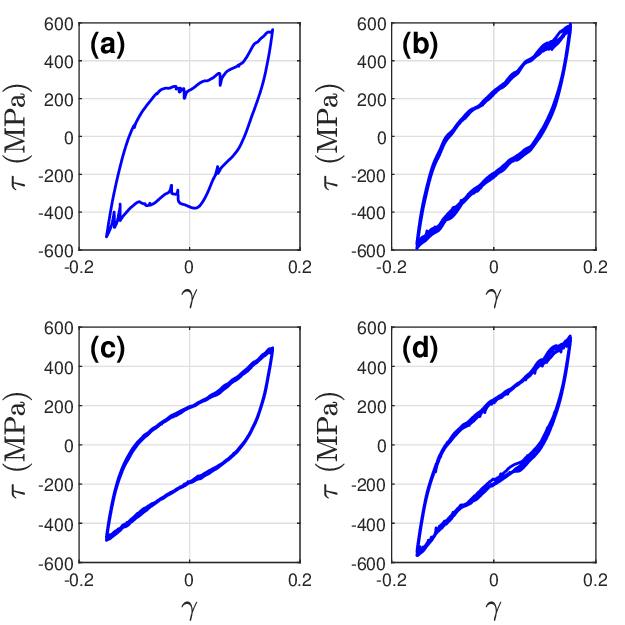}
    \caption{Shear stress-strain response for various cases of number of particles and polydispersity, over five cycles. The values of the parameters and the nomenclature used for the subfigures are listed in Tab.~\ref{tab:A4_check_part_num_size}.}
    \label{fig:A4_check_part_num_size}
\end{figure}

This appendix provides details on the conditions at which the mechanical response of the representative elementary volume becomes independent on the number, the size and the polydispersity of particles, in order to get ride of the discrete and crystallized nature of the granular sample. The nominal configuration described in this study is a cubic volume containing $N_p=10^3$ particles with mean radius $r_p=1$~mm and $dr_p/r_p=25$\% polydispersity.
To justify this choice, we simulate and compare the stress-strain relation at the nominal packing fraction and shear amplitude $(\phi,\gamma_m) = (0.70,0.15)$ for the few different configurations $(N_p,r_p,dr_p/r_p)$ summarized in Tab.~\ref{tab:A4_check_part_num_size}. Note that the representative elementary volume is kept constant in this parametric study, $N_pr_p^{3}=\mbox{const}$, to ensure that similar stress-strain curves rely on regimes with comparable energies. The four corresponding stress-strain responses are presented in Fig.~\ref{fig:A4_check_part_num_size}.
Looking first at the effect of the number of particles, the comparison of Figs.~\ref{fig:A4_check_part_num_size}a and~\ref{fig:A4_check_part_num_size}b shows that $N_p = 5^3$ generates a non-smooth stress-strain relation as a hallmark of the discrete nature of sudden and local stick-slip rearrangements between a small number of frictional contacts, while eight times more particles, $N_p = 10^3$, smooths the response by better averaging over more particles and events. However, the comparison between Figs.~\ref{fig:A4_check_part_num_size}b and~\ref{fig:A4_check_part_num_size}c shows that increasing the number of particles by another factor eight beyond the nominal configuration does not lead to further improvement. However, it is worth noting that this latter case is part of the dataset, see Tab.~\ref{tab:input_parameters} and Figs.~\ref{fig:4_parametric_results},~\ref{fig:5_parametric_analysis},~\ref{fig:6_partition_function} and ~\ref{fig:7_energies_loss_factor}.
Looking then at the effect of the polydispersity, the quantitative and qualitative similarities of the responses shown in Figs.~\ref{fig:A4_check_part_num_size}b and~\ref{fig:A4_check_part_num_size}d indicates that increasing size randomness beyond the nominal case does not bring significant differences: we thus infer that the sample is already amorphous enough (i.e. its crystallization is reliably prevented) with a $25\%$ polydispersity.
With a max-to-min size ratio of $1.25/0.75\simeq1.66$, our size dispersion is indeed slightly larger than the bidispersity considered in~\cite{Makse2004} (a distribution of grain radii in which $R_1=0.105$~mm for half the grains and $R_2=0.095$~mm for the other half; size ratio of $1.1$), the one in~\cite{Xing2021} (an ensemble of beads with $5$~mm and $6$~mm diameters; size ratio of $1.2$), and the one in~\cite{OHern2003} ($50$\---$50$ mixtures of particles with a size ratio of $1.4$), for the same purpose.

\end{appendix}


\section*{Declarations}

\subsection*{Authors’ Contributions}
S.J. conducted the investigations. A.F. and S.J. did the investigations, prepared the figures, wrote and reviewed the manuscript. J.B., N.P. and  J.L.D. discussed the results and reviewed the manuscript.

\subsection*{Data availability}
The authors declare that the data supporting the findings of this study are available within the paper itself. Raw data and computer codes are available from the corresponding author on reasonable request.

\subsection*{Conflict of interest}
The authors declare that they have no conflicts of interest with this work. They do not have any commercial or associative interest that represents a conflict of interest in connection with the work submitted. The work described has not been submitted elsewhere for publication, in whole or in part, and all the authors listed have approved the manuscript that is enclosed.

\subsection*{Funding}
This study is based on the doctoral researches of the first author (A.F.) which were fully and equally funded by both our institutions, ISAE-Supméca and Estaca.

\subsection*{Ethics approval and consent to participate}
Not applicable


\bibliographystyle{unsrtnat}

\end{document}